\documentclass[11pt]{article}

\usepackage{lineno,hyperref}
\usepackage{amsfonts,amsmath,times,amsthm,amssymb}
\usepackage[linesnumbered,algoruled,boxed,lined]{algorithm2e}
\usepackage{tikz}
\usepackage{multirow}
\usepackage{epstopdf}

\newtheorem{mydef}{Definition}

\newcommand{\vA}{\mathbf{A}}
\newcommand{\vD}{\mathbf{D}}
\newcommand{\vL}{\mathbf{L}}
\newcommand{\vEll}{\boldsymbol{\mathcal{L}}}
\newcommand{\vI}{\mathbf{I}}
\newcommand{\Vol}{{\rm Vol}}
\newcommand{\Cut}{{\rm Cut}}
\newcommand{\vc}{\mathbf{c}}
\newcommand{\Ncut}{{\rm Ncut}}
\newcommand{\Tr}{{\rm Tr}}
\newcommand{\vP}{\mathbf{P}}
\newcommand{\vC}{\mathbf{C}}
\newcommand{\vv}{\mathbf{v}}
\newcommand{\vs}{\mathbf{s}}
\newcommand{\vB}{\mathbf{B}}
\newcommand{\vone}{{\bf 1}}
\newcommand{\ER}{Erd\"{o}s-R\'{e}nyi}
\newcommand{\E}{\mathbb{E}}
\newcommand{\Var}{\mathbb{V}{\rm ar}}
\newcommand{\veps}{\mathbf{\epsilon}}

\begin{document}

\title{ \huge Clique-based Method for Social Network Clustering}

\author{ Guang Ouyang, 
       Dipak K. Dey, 
       Panpan Zhang} 


\maketitle

\begin{abstract}
In this article, we develop a clique-based method for social network clustering. We introduce a new index to evaluate the quality of clustering results, and propose an efficient algorithm based on recursive bipartition to maximize an objective function of the proposed index. The optimization problem is NP-hard, so we approximate the semi-optimal solution via an implicitly restarted Lanczos method. One of the advantages of our algorithm is that the proposed index of each community in the clustering result is guaranteed to be higher than some predetermined threshold, $p$, which is completely controlled by users. We also account for the situation that $p$ is unknown. A statistical procedure of controlling both under-clustering and over-clustering errors simultaneously is carried out to select localized threshold for each subnetwork, such that the community detection accuracy is optimized. Accordingly, we propose a localized clustering algorithm based on binary tree structure. Finally, we exploit the stochastic blockmodels to conduct simulation studies and demonstrate the accuracy and efficiency of our algorithms, both numerically and graphically.
\end{abstract}


\section{Introduction}
\label{Sec:Intro}
Networks are proliferating all around us, and they appear in different forms, such as (hardwired) electrical grids or (virtual) social relationships. Networked systems spread in various scientific and applied disciplines, for example, the Internet, the World Wide Web, metabolic networks, neural networks, food webs and social networks, etc. In this paper, we place our focus on social networks. Social network analysis is a branch of the social science which is an academic discipline studying a society and the behavior of entities therein. In sociometric or other quantitative studies, social networks are usually modeled by graph structures consisting of a set of nodes or vertices connected by directed arcs or undirected edges. More specifically, the {\em actors} in a social network are represented by nodes. For each pair of nodes, if there appears some pattern of {\em ties}, they will be connected by an edge. Edges can be directed or undirected, depending on the feature and interpretation of the network. There are a variety of social networks, such as the Facebook~\cite{Wohlgemuth}, marriage networks~\cite{Pei} and smart business networks~\cite{Pao}.

Due to solid theoretical foundation of statistics, statistical methodologies have been a significant input to the studies of social networks. A plethora of statistical methods have been established and developed to uncover relational structure of social networks, and dyadic ties between actors therein. In this paper, we focus on a significant research topic in social networks---{\em clustering}. More precisely, our goal is to develop some statistical methods for accurately clustering actors in a social network into mutually exclusive communities. This process, in many literatures, is called ``community detection.'' 

Roughly speaking, the fundamental principle underlying social network clustering is that a group of actors who are excessively connected are more likely to form a community. Technically speaking, the formation of a community requires that the connections of actors within the community are significantly higher than the connections between actors from different communities. From the sociological point of view, the occurrence of high connection density between actors in a community is usually due to some kind of homology or homogeneity of actors. For instance, consider a friendship network on Facebook. Intrinsically, students from the same department of a college are more likely form a friendship community, as they have a very high probability to know and friend each other. The homogeneities are reflected in the location parameter (i.e., college) and the academic parameter (i.e., department). On the other hand, students with different education, social or geographic background are much less likely to be connected. 

Past research on social network clustering can be summarized into two categories. One approach is to propose a (parametric or nonparametric) probabilistic graphical model (PGM) which characterizes the community structures of a social network. Pioneering work of such model-based approach were the $p_1$ model~\cite{Holland1981} and the stochastic blockmodel (SBM)~\cite{Holland1983}. Several successful models were proposed for community detection in the last two decades, including an extension of the SBM~\cite{Snijders}, a latent position model~\cite{Hoff}, a latent position cluster model~\cite{Handcock}, and a mixed membership SBM in~\cite{Airoldi}, etc. We refer the interested readers to~\cite{Goldenberg} for a complete and comprehensive review of PGMs for social network clustering. Another approach is to consider a metric that can be used to quantitatively evaluate the quality of social network clustering. The task is to specify an objective function based on the proposed metric, and the ultimate goal is to design an efficient algorithm to optimize the objective function over all potential clustering strategies. In this article, we call this class of approaches {\em metric-based methods}.  Precursory work in this direction traced back to~\cite{Watts}, in which a measure called cluster coefficient was proposed to evaluate mutual acquaintance between actors in a social network. We refer the interested readers to~\cite{Newman2001a} and~\cite{Newman2001b} for extensive discussions about cluster coefficient. While cluster coefficient is the metric that is often used for random graphs or dynamic network models, we place our focus on clustering problems of static networks in this manuscript. Representative and popular methods for static network clustering are the spectral clustering method~\cite{Ng}, the normalized cut approach~\cite{Shi}, and the modularity maximization method~\cite{Newman2006}, etc.

The contribution of this paper is as follows:
\begin{enumerate}
	\item We propose a novel clustering method based on graph cliques such that the clique score (a new index defined to evaluate clustering outcomes) of each identified community is higher than some threshold $p$.
	\item When threshold $p$ is unknown, we develop a systematic strategy to select localized thresholds $p$ (for each subnetwork which requires further subdivision) by controlling over-clustering and under-clustering simultaneously.
	\item The two algorithms proposed in this manuscript are efficient, easy to implement, and provide consistently reliable clustering results.
\end{enumerate}

We would like to point out that even though undirected binary social networks are considered in this manuscript for the sake of interpretation, our method can be extended to directed or weighted networks in a similar manner, done mutatis mutandis. The rest of this manuscript is organized as follows. In Section~\ref{Sec:notations}, we introduce some notations that will be used throughout the paper. In Section~\ref{Sec:classic}, we briefly review two classical clustering methods. Section~\ref{Sec:novel} is divided into two subsections. In Section~\ref{Sec:clique}, we propose a new metric to quantitatively measure the quality of clustering results, and then establish an objective function based on it, followed by an algorithm which is designed upon the idea of recursive bipartition to optimize the objective function. In Section~\ref{Sec:localized}, we consider the situation that the global parameter $p$ is not predetermined, and develop a method to compute localized parameter $p$ so as to maximize the detection accuracy of our algorithm. We further propose a localized algorithm (corresponding to localized parameter $p$), with some modifications on the algorithm developed in Section~\ref{Sec:clique}. We present several simulation examples in Section~\ref{Sec:simulation} to demonstrate the efficiency of our algorithms and accuracy and consistency of clustering results. In addition, we numerically compare the performance of the proposed algorithm and the traditional modularity maximization algorithm. Lastly, we address some concluding remarks and potential future studies in Section~\ref{Sec:con}.

\section{Notations}
\label{Sec:notations}
In this section, we introduce the notations that will be used throughout the manuscript. Let $G = G(V, \vA)$ be an undirected graph to model a social network of  $|V| = n$ actors, where $\vA$ is an $n \times n$ {\em adjacency matrix} such that
$$\vA = (a_{ij})_{n \times n},$$
in which 
$$a_{ij} = \begin{cases}
1, &\qquad \mbox{Nodes $i$ and $j$ are connected},
\\
0, &\qquad \mbox{otherwise},
\end{cases}$$
for $i, j \in V$. Let $\vD$ be an $n \times n$ diagonal matrix such that 
$$\vD = \begin{pmatrix}
\deg{(1)} & 0 & \cdots & 0
\\ 0 & \deg{(2)} & \cdots & 0
\\ \vdots & \vdots & \ddots & \vdots
\\ 0 & 0 & \cdots & \deg{(n)}
\end{pmatrix},$$
where $\deg{(i)}$ represents the {\em degree} of node $i \in V$. Consider the matrix $\vL$ defined as
$$\vL = \vD - \vA.$$
The {\em normalized Laplacian matrix} of $G$~\cite[page 2]{Chung} is then given by
$$\vEll = \vD^{-1/2} \vL \vD^{-1/2} = \vI - \vD^{-1/2} \vA \vD^{-1/2},$$
where $\vI$ denotes the {\em identity matrix} of rank $n$.

Two graph invariants on which many classical network clustering methods depend on are {\em volume} and {\em cut}. The volume of a graph $G$, denoted $\Vol(G)$, is the total number of degrees of the nodes in $G$, i.e., 
$$\Vol(G) = \sum_{i = 1}^{n} \deg{(i)}.$$ 
The cut, on the other hand, is defined on subgraphs of $G$. Let $G_1$ and $G_2$ be two disjoint subgraphs of $G$, the cut of $G_1$ and $G_2$, denoted $\Cut(G_1, G_2)$, is the number of edges linking $G_1$ and $G_2$, i.e.,
$$\Cut(G_1, G_2) = \sum_{i \in G_1, j \in G_2} a_{ij}.$$ 

\section{Prerequisites}
\label{Sec:classic}
In this section, we give brief reviews of the spectral clustering method with a concentration on the normalized cut and the modularity maximization approach as prerequisites. 
\subsection{Spectral network clustering}
\label{Sec:spectral}
We first look at the {\em spectral network clustering algorithm} developed in~\cite{Shi} and~\cite{Ng}, inspired from the spectral graph theory~\cite{Chung}. Generally speaking, a measure called the {\em normalized cut} was adopted as an objective function to quantify the number of edges across different communities. Suppose that a social network $G$ is split into $h$ communities $G_1, G_2, \ldots, G_h$. The normalized cut is given by
\begin{equation}
	\label{Eq:ncut}
	\Ncut(G_1, G_2, \ldots, G_h) = \sum_{k = 1}^{h} \frac{\Cut(G_k, \overline{G_k})}{\Vol(G_k)},
\end{equation}
where $\overline{G_k}$ denotes the complement of $G_k$ in $G$, for $k = 1, 2, \ldots, h$. Consider an $n \times h$ community indicating matrix $\vP = (p_{ik})_{n \times h}$, in which each entry is $p_{ik} = 1/\sqrt{\Vol(G_k)}$ for node $i$ in community $G_k$. The optimization problem corresponding to Equation~(\ref{Eq:ncut}) is, in fact, a discrete {\em trace minimization} problem, which can be approximated by a standard trace minimization problem with a relaxation of the discreteness condition as follows:
\begin{equation}
	\label{Eq:ncutrelax}
	\min_{G_1, G_2, \ldots, G_h} \Tr(\vP^{\top} \vL \vP) \approx \min_{\vP \in \mathbb{R}^{n \times h}} \Tr(\vP^{\top} \vL \vP),
\end{equation}
subject to $\vP^{\top} \vD \vP = \vI$. Let $\vC = \vD^{1/2} \vP$. The optimization problem on the right hand side of Equation~(\ref{Eq:ncutrelax}) is equivalent to
$$\min_{\vC \in \mathbb{R}^{n \times h}} \Tr(\vC^{\top} \vEll \vC),$$
subject to $\vC^{\top} \vC = \vI$. This is a standard {\em Rayleigh quotient} problem, and the solution of $\vC$ is composed of the first $h$ eigenvectors of $\vEll$; see~\cite{Chung} and~\cite{Horn}. Thus, the solution to the right hand side of Equation~(\ref{Eq:ncutrelax}) can be obtained by solving the first $h$ eigenvectors in a generalized eigenvalue system $\vL \vv = \lambda \vD \vv$, which can be done via an adaptive algorithm based on {\em bipartition} proposed in~\cite{Shi}.

\subsection{Modularity maximization}
\label{Sec:modularity}
Another popular algorithm for social network clustering, known as the {\em modularity maximization}, was first introduced in~\cite{Newman2006}. In essence, the underlying principle is to partition a social network into mutually exclusive communities such that the number of edges across different communities is significantly less than the expectation, whereas the number of edges within each community is significantly greater than the expectation. In~\cite{Newman2006}, a bipartition situation was considered, and the clustering outcome was evaluated by a measure---{\em modularity}---defined as
$$Q = \frac{1}{2 \Vol(G)} \vs^{\top} \vB \vs,$$
where $\vs$ is an $n \times 1$ column indication vector such that
$$\begin{cases}
s_i = 1, &\qquad\mbox{if node $i$ belongs to community 1};
\\ s_i = -1, &\qquad\mbox{if node $i$ belongs to community 2},
\end{cases}$$
and $\vB = (b_{ij})_{n \times n}$ is an $n \times n$ matrix with entires
$$b_{ij} = a_{ij} - \frac{\deg{(i)} \deg{(i)}}{\Vol{(G)}},$$
for $i = 1, 2, \ldots, n$. Subdivisions for existing communities are available by repeatedly implementing the proposed modularity algorithm. The decision of whether or not subdividing an existing community $G_{k}$ of size $n^{G_k}$ depends on an associated {\em modularity matrix} $\vB^{G_k} = \left(b^{G_k}_{ij}\right)_{n^{G_k} \times n^{G_k}}$ with entries
$$b^{G_k} _{ij} = b_{ij} - \delta(i, j) \sum_{l \in G_k} b_{il},$$
where $\delta(\cdot, \cdot)$ denotes the {\em Kronecker delta function}. Subdivision for $G_k$ is terminated if the largest eigenvalue of $\vB^{G_k}$ is zero. 

\section{Clique-based clustering algorithm}
\label{Sec:novel}
In this section, we propose a novel algorithm for social network clustering. More specifically, we define a new measure called the {\em $p$-clique index} to quantitatively evaluate the quality of clustering outcome. The section is divided into two subsections. We introduce a clique-based clustering algorithm in Section~\ref{Sec:clique}, and then extend it to a localized clique-based clustering algorithm in Section~\ref{Sec:localized}. The localized algorithm enables us to update the threshold $p$ for each subnetwork so as to well control the potential over-clustering or under-clustering problems. We will discuss the details in the sequel.

\subsection{Clique-based clustering algorithm}
\label{Sec:clique}
In this section, We give a clique-based clustering algorithm, inspired from the modularity maximization algorithm proposed in~\cite{Newman2006}. In graph theory, a {\em clique} is defined as a complete graph on a set of nodes, i.e., each pair of nodes is connected by an edge. A clique is called {\em maximal} if it cannot be extended to any larger-size clique by including any adjacent node. The ideal clustering outcome is that each community in a network is a maximal clique. In reality, this is hard to achieve. For many real world social networks, it is even difficult to guarantee that each community forms a clique. Therefore, an appropriate measure to assess the degree of connectivity of a community is needed. The next graph invariant measure can be used to gauge the internal link density of each community in a network.
\begin{mydef}[clique score]
	The clique score of a community (cluster, subnetwork) is the ratio of the number of observed ties to the number of edges in the clique over the same number of nodes.
\end{mydef}
For example, a community consisting of 10 nodes and 18 internal links has clique score $18/\binom{10}{2} = 0.4$. It is obvious that a clique always has clique score one. Our goal is to develop an adaptive algorithm to maximize the overall clique scores (e.g., weighted average) of all communities in a network, subject to the clique score of each community exceeding some predetermined threshold $0 \le p \le 1$. The condition of ``exceedance'' in our algorithm is essential as it guarantees that none of the communities in our clustering outcome performs extremely bad or does not achieve the minimum standard. In addition to this, the choice of $p$ is flexible, depending on the users' needs or the realistic features of communities in a social network. In Section~\ref{Sec:localized}, we will discuss how to choose an appropriate value of $p$ to optimize the performance of the proposed algorithm when no prior information of $p$ is available. In the next definition, we bridge the gap between clique graph and $p$.
\begin{mydef}[$p$-clique]
	\label{Def:ER}
	A $p$-clique is a random graph of a set of nodes, of which each pair is connected by an edge independently with probability $p$, for $0 \le p \le 1$.
\end{mydef}
The $p$-clique defined in this manuscript is not novel, and structurally it is equivalent to the {\em \ER\ model} proposed by~\cite{Erdos}. The definition of $p$-clique (c.f.\ Definition~\ref{Def:ER}) adopts an alternative interpretation of the \ER\ model given by~\cite{Gilbert}. Since the nodes are connected independently, there is no structure of communities or clusters in $p$-cliques theoretically. Hence, $p$-cliques appear to be a proper benchmark model in our study. The expected number of edges in a $p$-clique on $n$ nodes is $np(1 - p)$.  

Suppose that a social network $G$ of size $n$ is clustered into $h$ communities, $G_1, G_2, \ldots, G_h$ respectively with community sizes $n_1, n_2, \ldots, n_h$ such that $\sum_{k = 1}^{h} n_k = n$. Our task is to search for a clustering rule such that the total degree of nodes in each community is significantly larger than the expected number of edges of a $p$-clique of the same size, whereas the total number of links across different communities is minimized. Based on this idea, we propose a measure called the {\em $p$-clique index} as follows, and our ultimate goal is to design an adaptive algorithm for an optimal clustering rule where the $p$-clique index is maximized.
\begin{mydef}[$p$-clique index]
	\label{Def:pclique}
	Let $G$ be a social network that consists of $n$ nodes, and $C = [G_1, G_2, \ldots, G_h]$ be a clustering rule which divides $G$ into $h$ communities. The $p$-clique index is given by
	\begin{align}
		D(\vc, p) &= \frac{1}{n(n - 1)} \left(\sum_{k = 1}^{h} \bigl(\Vol(G_k) - p n_k (n_k - 1)\bigr) \right. \nonumber
		\\ &\qquad{}- \left. \sum_{1 \le k \neq l \le h} \bigl({\rm Cut}(G_k, G_l) - p n_k n_l \bigr) \nonumber \right) 
		\\ &= \frac{1}{n(n - 1)} \sum_{1 \le i \neq j \le n} \bigl((a_{ij} - p)\delta(c_i, c_j) + (p - a_{ij})(1 - \delta(c_i, c_j))\bigr),
		\label{Eq:pindex}
	\end{align}
	where $\vc = (c_1, c_2, \ldots, c_n)^{\top}$ is the membership indication vector for nodes.
\end{mydef}
The essence of $p$-clique index is to reward connected nodes in the same community (with $1 - p$) and disconnected nodes from different communities (with $p$), but penalize connected nodes from different communities (with $-(1 - p)$) and disconnected nodes in the same community (with $-p$). Different from model-based methods, this approach is not to fit data (i.e., observation) to any type of generative model or $p$-clique structures. Rather, our goal is to determine a clustering rule, in which each community has a higher clique score than a predetermined threshold $p$.

The algorithm for our clique-based clustering approach is based on the hierarchical clustering algorithm developed for modularity maximization in~\cite{Newman2006}. To begin with, we consider bipartition, i.e., clustering a social network into two communities $1$ and $2$. Define an alternative membership indication vector, $\vs = (s_1, s_2, \ldots, s_n)^{\top}$, as follows:
$$\begin{cases}
s_i = 1, \qquad &\mbox{if node $i$ belongs to community $1$},
\\ s_i = -1, \qquad &\mbox{if node $i$ belongs to community $2$},
\end{cases}$$
for $i = 1, 2, \ldots, n$. Then, the $p$-clique index (c.f.\ Equation~(\ref{Eq:pindex})) is equivalent to 
$$
D(\vs, p) = \frac{1}{n(n - 1)} \sum_{1 \le i \neq j \le n} (a_{ij} - p) s_i s_j = \frac{1}{n(n - 1)} \vs^{T} \vC(p) \vs,
$$
where the {\em $p$-clique matrix} $\vC(p)$ is given by
$$\vC(p) = \vA - p(\vone_{n \times n} - \vI_{n \times n}),$$
in which $\vone_{n \times n}$ is an $n \times n$ matrix of all ones. In what follows, the network clustering is converted to an optimization problem
\begin{equation}
	\label{Eq:opti1}
	\max_{\vs} \left\{\vs^{\top} \vC(p) \vs\right\}.
\end{equation}
This is equivalent to
\begin{align}
	&\quad{}\max_{\vs} \left\{D(\vs, p) - D(\vone_{n \times 1}, p)\right\} \nonumber
	\\ &= \frac{1}{n(n - 1)}\max_{\vs} \left\{ \vs^{\top} \vC(p) \vs - \vone^{\top}_{n \times 1} \vC(p) \vone_{n \times 1}\right\} \nonumber
	\\ &= \frac{1}{n(n - 1)}\max_{\vs} \left\{ \vs^{\top} \left( \vC(p) - \left(\frac{1}{n} \sum_{i, j = 1}^{n} C(p)_{ij} \right) \vI_{n \times n} \right) \vs\right\}, \label{Eq:opti2}
\end{align}
where $C(p)_{ij}$ denotes the $(i, j)$th entry of $p$-clique matrix $\vC(p)$. Since $\vs = \{-1, 1\}^n$ is dyadic, the optimization problem is NP-hard. To relax the problem, our strategy is to allow $\vs$ to be any normalized real-valued vector. One solution to Equation~(\ref{Eq:opti2}) is the eigenvector $\vv = (v_1, v_2, \ldots, v_n)^{\top}$ corresponding to the largest eigenvalue of $\vC(p) - \left(\frac{1}{n} \sum_{i, j = 1}^{n} C(p)_{ij} \right) \vI_{n \times n}$, denoted by $\lambda_{\rm max}$. We thus obtain a natural approximation of the solution; that is, we cluster the nodes with respect to the signs of the components of $\vv$:
$$\begin{cases}
s^{\rm approx}_i = 1, \qquad &v_i \ge 0,
\\ s_i^{\rm approx} = -1, \qquad &v_i < 0,
\end{cases}$$
for $i = 1, 2, \ldots, n$.

We remark that we would rather work on the optimization problem in Equation~(\ref{Eq:opti2}) than that in Equation~(\ref{Eq:opti1}) analytically, albeit they are mathematically identical. The underlying reason is that Equation~(\ref{Eq:opti2}) provides an instinctive insight of the decision rule of partitioning an existing community. Our rule is that we do not further subdivide the existing cluster if the eigenvalue $\lambda_{\rm max} \le 0$ and $D(\vone_{n \times 1}, p) \ge 0$. Both of the conditions are needed since $\lambda_{\rm max} \le 0$ indicates that $D(\vone_{n \times 1}, p) \ge \max\{\max_{\vs} \{D(\vs, p)\}, 0 \}$ and $D(\vone_{n \times 1}, p) \ge 0$ implies that the link density of the existing community is already higher than $p$.

We next propose an algorithm based on recursive bipartition for our $p$-clique approach. In practice, we first determine the $p$-clique matrix, $\vC^{(G)}(p)$ for a given network $G$ of size $n$. We partition $G$ into two clusters $G_1$ and $G_2$ with respect to the signs of the entries of the eigenvector corresponding to the largest eigenvalue of $\vC^{(G)}(p)$, and then compute the additional contribution to the $p$-clique index due to the division, denoted by 
$$\Delta D(p) = \frac{1}{n(n - 1)} \sum_{i, j \in G} \left(s_i C^{(G)}(p)_{ij} s_j - C^{(G)}(p)_{ij} \right).$$
We continue to apply the bipartition algorithm respectively to $G_1$ (according to the associated $p$-clique matrix $\vC^{(G_1)}(p)$, where $\vC^{(G_1)}(p)$ is a submatrix extracted directly from $\vC^{(G)}(p)$ only for the nodes contained in $G_1$) and $G_2$ (according to the associated $p$-clique matrix $\vC^{(G_2)}(p)$ obtained in a similar manner), if at least one of the two regularity criteria is met. We terminate the algorithm until no further partition is needed. The algorithmic procedure is presented in Algorithm~\ref{Alg:bipar}.
\vspace{2.5mm}

\IncMargin{1em}
\begin{algorithm}[H]
	\KwIn{$p$-clique matrix $\vC^{(G)}(p)$ of network $G$, empty community list $L$} 
	\ProcSty{\ProcFnt Procedure} ${\rm BiPartition}(G, \vC^{(G)}(p))$\;
	Compute the eigenvector $\vs$ corresponding to the largest eigenvalue of $\vC^{(G)}(p)$\;
	Partition $G$ into $G_1$ and $G_2$ according to the signs of entries of $\vs$\;
	Compute the additional contribution to $p$-clique index, $\Delta D(p)$\;
	\uIf{$\Delta D(p) > 0$ {\rm or} $\sum_{i,j \in G} C^{(G)}_{ij}(p) < 0$}{${\rm BiPartition}(G_1, \vC^{(G_1)}(p))$\;
		${\rm BiPartition}(G_2, \vC^{(G_2)}(p))$\;}
	\Else{Add $G$ to community list $L$}
	\KwOut{Community list $L$}
	\caption{The bipartition-based algorithm for $p$-clique index maximization}
	\label{Alg:bipar}
\end{algorithm}
\DecMargin{1em}
\vspace{2.5mm}

To conclude this section, we address some remarks on the proposed algorithm. The essence of Algorithm~\ref{Alg:bipar} is to maximize the additional contributions to the overall $p$-clique index, whenever a division of an existing cluster is implemented (versus its current state) . As mentioned, the optimization problem is NP-hard, and our strategy is to relax the problem by allowing $\vs$ to be any real vector with $\ell_2$ norm equal to one. Our goal is to determine the eigenvector associated with the largest eigenvalue of the $p$-clique matrix, and specifically our strategy is to exploit an {\em implicitly restarted Lanczos method} developed in~\cite{Calvetti}. It is worthy of noting that our algorithm will continue to be executed if $\sum_{i,j \in G^{\prime}} C^{(G^{\prime})}_{ij}(p) < 0$ for some $G^{\prime}$ even when $\Delta D(p) \le 0$, since, under such circumstance, the clique index of $G^{\prime}$ is less than $p$. We take the quality (i.e., having clique score higher than predetermined threshold $p$) of each resulting community as a higher priority, bearing with non-optimal grade for the overall $p$-clique index. We view this as a major feature, as well as an advantage, of our algorithm, since the choice of threshold $p$ is absolutely flexible, depending on the users' needs. 

On the other hand, a natural question arising from our algorithm is `` what if there is no specific requirement or prior information for $p$?'' An arbitrary choice of $p$ may lead to over-clustering or under-clustering problems. Having this concern in mind, we propose a modified algorithm based on Algorithm~\ref{Alg:bipar} in Section~\ref{Sec:localized}.In addition, we carry out a statistical procedure for ``reasonable'' selection of unspecified $p$ so as to simultaneously minimize the errors of over-clustering and under-clustering.  

\subsection{Localized clustering algorithm}
\label{Sec:localized}
We have demonstrated that the algorithm proposed in Section~\ref{Sec:clique} is advantageous for social network clustering, as it allows users to control the quality of each resulting community. In many cases, the threshold parameter $p$ is predetermined or preselected, based on users' needs or past research experience. For the cases that $p$ is not specified, it is needed to develop a systematic method to choose an optimal assignment of $p$. Probably a question that needs to be answered in advance is whether or not there exists such optimum $p$. An analogous issue occurs for the modularity maximization algorithm, and was discussed in~\cite{Reichardt} and~\cite{Lancichinetti}, where the conclusion from both was negative for the existence of such overall optimal value of $p$. In this section, we consider a localized clustering strategy, the core idea of which is to select different values of $p$ for different subnetworks. 

In essence, our strategy for the selection of $p$ is a process of balancing over-clustering and under-clustering. Intrinsically, a large value of $p$ usually results in small sizes of clusters (i.e., over-clustering), whereas a small value of $p$ sometimes fails to guarantee the quality of clusters (i.e., under-clustering). The procedure that balances between over-clustering and under-clustering is analogous to dealing with type {\rm I} errors and type {\rm II} errors in statistical hypothesis testings. We borrow these two terminologies in our study. Let us call the error of over-clustering as type {\rm I} error and the error of under-clustering as type {\rm II} error. Our goal is determine a distinct value of $p$ for each existing cluster such that both types of errors are well controlled.

Recall the $p$-clique (i.e., the \ER\ model) introduced in Definition~\ref{Def:ER}. The model is constructed completely at random without any cluster structure, suggesting that we can exploit it to control type {\rm I} error, and accordingly determine an upper bound of $p$. Consider an \ER\ graph, ${\rm ER}(n, p_0)$, where $n$ and $p_0$ are given. Intuitively, we need to set the threshold parameter $p$ significantly less than $p_0$ so that there is a small probability to divide ${\rm ER}(n, p_0)$ into two subgraphs or more by our algorithm. Assume that $\alpha$ is the maximal percentage of the number of nodes to be split from the \ER\ graph under tolerance. Our goal turns to find the largest value of $p$ such that at most $\alpha n$ nodes are split off of the majority. Let $X$ denote the observed inter-across link density between the split group (i.e., $\alpha n$ nodes) and the remainder (i.e., $(1 - \alpha)n$ nodes), and $F_X$ be its distribution. An instinctive and reasonable choice of $p$ is thus the $\alpha$th percentile of $F_X$. Although the expected number of inter-across edges is $(1 - \alpha)np_0$ in theory, the inter-across link density is anticipated to be much smaller than the internal density within the large group of $(1 - \alpha) n$ nodes. In this regard, we approximate $F_X$ by a truncated normal distribution, where the top $\alpha$ is curtailed; that is, $X$ is normally distributed with mean
\begin{equation*}
	\label{Eq:mean}
	\E[X] = p_0 - \frac{\phi(z_\alpha)}{1 - \alpha} \sqrt{\frac{p_0(1 - p_0)}{(1 - \alpha)n}},
\end{equation*}
and variance
\begin{equation*}
	\Var[X] = \frac{p_0 (1 - p_0)}{(1 - \alpha)n} \left[1 - \frac{z_{\alpha} \phi(z_{\alpha})}{1 - \alpha} - \left(\frac{\phi(z_\alpha)}{1 - \alpha}\right)^2\right],
\end{equation*}
where $\phi(\cdot)$ denote the density function of the standard normal distribution, and $z_{\alpha}$ is the $(1 - \alpha)$th percentile of standard normal. Thus, we obtain an upper bound for $p$ as a function of $\alpha$, which is given by
\begin{align}
	p_{\rm U}(\alpha) &= \max \left\{0, \E[X] - z_\alpha \sqrt{\Var[X]} \right\} \nonumber 
	\\ &= \max \left\{0, p_0 - \xi(\alpha) \sqrt{\frac{p_0 (1 - p_0)}{(1 - \alpha)n}}\right\},
	\label{Eq:pupper}
\end{align}
where $\xi(\alpha)$ is a constant depending on $\alpha$ only.

The next task is to determine the lower bound for $p$. Suppose that we have two independent \ER\ graphs which are respectively denoted by ${\rm ER}(n_1, p_1)$ and ${\rm ER}(n_2, p_2)$, and $p_{12}$ is the link density between the two graphs. Let $\beta$ be the probability that the two graphs are merged. We once again consider normal approximation, and conclude that the two graphs are both statistically significant if we have
\begin{equation}
	\label{Eq:plower}
	\min\{p_1, p_2\} > p_{12} + z_{\beta}\sqrt{\frac{p_{12}(1 - p_{12})}{n_1 n_2}},
\end{equation}
the right-hand side of which in fact forms the lower bound for the parameter $p$. However, this lower bound is intractable in general as it requires the information of $p_{12}$, which is usually unknown for most real-world networks. Therefore, we place focus on the upper bound for $p$, and our strategy is to set the parameter $p$ as large as possible under tolerance so as to minimize the probability of merging significant subnetworks.

Finally, we propose a new algorithm for unspecified $p$ based on Algorithm~\ref{Alg:bipar} proposed in Section~\ref{Sec:clique}. The new algorithm is designed for controlling type {\rm I} and {\rm II} errors simultaneously. As mentioned, however, it seems to be unreasonable to have a global threshold $p$ for our algorithm. We elaborate the reasonings and demonstrate our solution via the following illustrative example. 

Suppose that we cluster a network $G$ of size $n$ into two subnetworks $G_1$ of size $n_1$ and $G_2$ of size $n_2$. Initially, we adopt a threshold $p^{(G)}$ which depends on the observed link density of $G$. However, if we continue to use $p^{(G)}$ as the threshold when $G_1$ or $G_2$ or both need to be further clustered, we may have to bear with over-clustering or under-clustering risks in the subdivision processes. Alternatively, we suggest to reset the threshold(s) before subdividing $G_1$ or $G_2$ or both. In other words, we choose
$$
p^{(G_1)} = \max\left\{0, p_1 - \xi(\alpha)\sqrt{\frac{p_1 (1 - p_1)}{(1 - \alpha)n_1}}\right\}$$
and
$$p^{(G_2)} = \max\left\{0, p_2 - \xi(\alpha)\sqrt{\frac{p_2 (1 - p_2)}{(1 - \alpha)n_2}}\right\}
$$
as updated thresholds for $G_1$ and $G_2$, respectively. From then on, the threshold parameters are refreshed in this manner for all subnetworks that need to be further subdivided. As for each subnetwork to which a fresh threshold parameter is assigned, we call the new algorithm {\em localized clustering algorithm}. 

One of the most effective ways to illustrate this heirachical clustering process is probably to exploit {\em binary tree}. We start with a root node that represents the original network which needs to be clustered. At the first level, two child nodes are attached to the root node. The child nodes represent either subnetworks or communities. If a child node is a subnetwork which requires further subdivision, it will carry over two higher-level child nodes and itself turns to an internal node. If a child node is a community which does not need further subdivision, it becomes a terminal node in the tree. We use rectangle for internal nodes and circle for terminal nodes. When the algorithm is terminated, the clustering result is reflected in all the terminal nodes in the binary tree. An example of binary tree is given in Figure~\ref{Fig:binary}.
\begin{center}
	\begin{figure}[ht]
		\centering
		\begin{tikzpicture}
		\node[rectangle, draw] at (0, 0) (G){Network $G$};
		\node[rectangle, draw] at (-3, -2) (G1){Subnetwork $G_1$};
		\node[rectangle, draw] at (3, -2) (G2){Subnetwork $G_2$};
		\node[circle, draw, scale = .5] at (-5, -4) (C1){Community 1};
		\node[rectangle, draw] at (-2, -4) (G3){Subnetwork $G_3$};
		\node[circle, draw, scale = .5] at (1, -4) (C2){Community 2};
		\node[circle, draw, scale = .5] at (5, -4) (C3){Community 3};
		\node[circle, draw, scale = .5] at (-4, -6) (C4){Community 4};
		\node[circle, draw, scale = .5] at (0, -6) (C5){Community 5};
		\draw (G)--(G1);
		\draw (G)--(G2);
		\draw (G1)--(C1);
		\draw (G1)--(G3);
		\draw (G2)--(C2);
		\draw (G2)--(C3);
		\draw (G3)--(C4);
		\draw (G3)--(C5);
		\end{tikzpicture}
		\caption{An example of binary tree that represents a result of five communities}
		\label{Fig:binary}
	\end{figure}
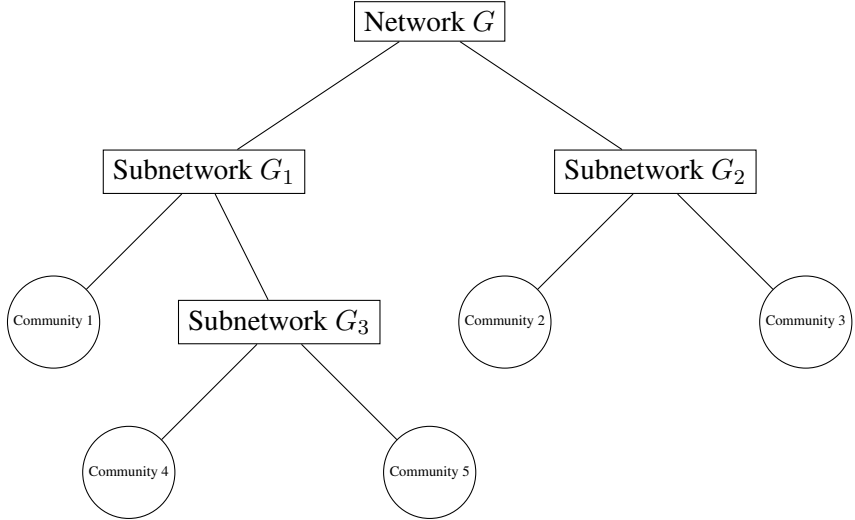
\end{center}

We are now in the position of defining a new clique index, the {\em local clique index}, modified from the $p$-clique index given in Definition~\ref{Def:pclique}.
\begin{mydef}[local clique index]
	Consider a network $G$ consisting of $n$ nodes. Let $T$ be the binary tree that describes a hierarchical clustering procedure on $G$, and $\{v\}$ be the collection of all internal nodes of $T$. The local clique index is given by
	\begin{equation}
		\label{Eq：lscore}
		{\rm LD}(T) = \frac{1}{n(n - 1)} \sum_{v \in T} \sum_{i \neq j \in v} \left((a_{ij} - p_v)\delta_{ij}^{(v)} + (p_v - a_{ij})\left(1 - \delta_{ij}^{(v)}\right)\right),
	\end{equation}
	where $p_v$ is the threshold of parameter $p$ for internal node (subnetwork) $v$.
\end{mydef}

As shown in Figure~\ref{Fig:binary}, our new algorithm is again based on recursive bipartition procedures, and the goal is to maximize the score function given in Equation~(\ref{Eq：lscore}). Analogous to Algorithm~\ref{Alg:bipar}, we need to determine the local clique matrix for each $v$ in $T$, i.e.,
$$\vC^{(v)} = \vA^{(v)} - p_v \left(\vone_{n_v \times n_v} - \vI_{n_v \times n_v} \right),$$
where $\vA^{(v)}$ is the updated adjacency matrix for subnetwork $v$ and $n_v$ is the size of $v$. Besides, we need to compute the additional contribution from each bipartition of $v$ to the overall score ${\rm LD}(T)$, i.e.,
$$\Delta {\rm LD}(v) = \frac{1}{n(n - 1)} \sum_{i \neq j \in v} \left((a_{ij} - p_v)\delta_{ij}^{(v)} + (p_v - a_{ij})\left(1 - \delta_{ij}^{(v)}\right)\right).$$
The algorithm for hierarchical clustering process is terminated as $\Delta {\rm LD}(v) \le \frac{1}{n(n - 1)} \sum_{i \neq j \in v} (a_{ij} - p_v)$ for all existing $v$ in $T$, and then all remaining internal nodes $v$ turn out to be terminal nodes, i.e., added to the community list. We summarize our strategy in Algorithm~\ref{Alg:localbipar}, slightly modified from Algorithm~\ref{Alg:bipar}.
\vspace{2.5mm}

\IncMargin{1em}
\begin{algorithm}[H]
	\KwIn{Network $G$, binary tree $T$ with a single root node representing $G$, tolerance of type {\rm I} error $\alpha$}
	\ProcSty{\ProcFnt Procedure} ${\rm LocalBiPartition}(G)$\;
	Determine network size $n_G$ and the clique score $p^{\rm obs}_G$ of $G$\;
	Compute the threshold $p_G = \max \left\{0, p^{\rm obs}_G - \xi(\alpha) \sqrt{\frac{p^{\rm obs}_G (1 - p^{\rm obs}_G)}{(1 - \alpha)n}}\right\}$\;
	Create local clique matrix $\vC^{(G)} = \vA^{(G)} - p_G \left(\vone_{n_G \times n_G} - \vI_{n_G \times n_G} \right)$\;
	Compute the eigenvector $\vs$ corresponding to the largest eigenvalue of $\vC^{(G)}$\;
	Partition $G$ into $G_1$ and $G_2$ with respect to the signs of $\vs$\;
	Compute additional contribution, $\Delta {\rm LD}(G)$, to overall local clique index by partitioning $G$\;
	\uIf{$\Delta {\rm LD}(G) > \frac{1}{n_G(n_G - 1)} \sum_{i \neq j \in v} (a_{ij} - p_G)$}{Add $G_1$ to the left child node position of $G$\; ${\rm LocalBiPartition}(G_1)$\; Add $G_2$ to the right child node position of $G$\;
		${\rm LocalBiPartition}(G_2)$\;}
	\KwOut{Binary tree $T$}
	\caption{The algorithm of network clustering based on local clique index maximization}
	\label{Alg:localbipar}
\end{algorithm}
\DecMargin{1em}
\vspace{2.5mm}

\section{Simulations}
\label{Sec:simulation}
In this section, we conduct some simulation studies to evaluate the performance of two clique-based clustering algorithms proposed in the previous sections. We also show that the proposed algorithm outperforms the modularity maximization approach~\cite{Newman2006} for network clustering. The SBM in~\cite{Snijders} is adopted to generate networks, as they allow us to predetermine the structure of simulated networks, which can be used as the ground truth for comparisons.

To begin with, we briefly review the SBM. Given a network of $n$ nodes which belong to $h$ nonempty communities, let $n_1, n_2, \ldots, n_h$ be respectively the size of each community. For $i = 1, 2, \ldots, n$, a mapping $c: \{1, 2, \ldots, n\} \mapsto \{1, 2, \ldots, h\}$ preserves the membership information for each node labeled with $i$. An $h \times h$ probability matrix $\vB$ describes the link densities within every community, as well as those between different communities, i.e., $\vB = \left(B_{c_{(i)}, c_{(j)}}\right)$. Notice that we have to require $B_{c_{(i)}, c_{(j)}} \le \min\left\{B_{c_{(i)}, c_{(i)}}, B_{c_{(j)}, c_{(j)}}\right\}$ for all $i, j$, when predefining matrix $\vB$; otherwise, the simulated network would be against the clustering structure obtained by our algorithm(s). Suppose that $\vB$ (the probabilistic structure of network) is specified, we are able to simulate the entries in the adjacency matrix 
$$a_{ij} = {\rm Bernoulli} \left(B_{c_{(i)}, c_{(j)}}\right)$$
for $i < j$. Suppose that the network is undirected, we have $a_{ji}$ equal $a_{ij}$ by symmetry. We also assume that  all the entries on the diagonal equal $0$, as loops are not considered in our study.

To quantitatively evaluate the performance of algorithms, we adopt two well-defined robust measures: the {\em normalized mutual information} (NMI) and the {\em Adjusted Rand Index} (ARI), respectively proposed in~\cite{Fred} and in~\cite{Hubert}. Suppose that $T$ is the ground truth of community structure, and $S$ is the clustering result of an algorithm, the NMI of $T$ and $S$ is given by
\begin{equation*}
	{\rm NMI}(T, S) = \frac{-2 \sum_{k = 1}^{C_T} \sum_{l = 1}^{C_S} N_{kl} \log \left(\frac{N_{kl} N_{\cdot \cdot}}{N_{k \cdot}N_{\cdot l}}\right)}{\sum_{k = 1}^{C_T} N_{k \cdot} \log\left(\frac{N_{k \cdot}}{N_{\cdot \cdot}}\right) + \sum_{l = 1}^{C_S} N_{\cdot l} \log \left(\frac{N_{\cdot l}}{N_{\cdot \cdot}}\right)},
\end{equation*}
where $C_T$ and $C_S$ are the number of communities for $T$ and $S$, respectively; $(N_{kl})$ is a $C_T \times C_L$ {\em confusion matrix}, in which $N_{kl}$ denotes the number of nodes that should be in community $k$ according to the truth, but are mis-clustered into community $l$ according to algorithm $S$; $N_{k \cdot}$, $N_{l \cdot}$, and $N_{\cdot \cdot}$ are standard definitions of the sum of the $k$th row, the sum of the $l$th column, and the overall sum of the confusion matrix, respectively. We borrow the same notations and give the definition of ARI as follows:

\begin{equation*}
	{\rm ARI}(T, S) = \frac{\sum_{k = 1}^{C_T} \sum_{l = 1}^{C_S} \binom{N_{kl}}{2} - \left[\sum_{k = 1}^{C_T} \binom{N_{k \cdot}}{2} \sum_{l = 1}^{C_S} \binom{N_{l \cdot}}{2}\right] \big{/} \binom{N_{\cdot \cdot}}{2}}{\frac{1}{2} \left[\sum_{k = 1}^{C_T} \binom{N_{k \cdot}}{2} + \sum_{l = 1}^{C_S} \binom{N_{l \cdot}}{2}\right] - \left[\sum_{k = 1}^{C_T} \binom{N_{k \cdot}}{2} \sum_{l = 1}^{C_S} \binom{N_{l \cdot}}{2}\right] \big{/} \binom{N_{\cdot \cdot}}{2}}.
\end{equation*}

In addition, we propose a measure analogous to NMI.  Once again, assuming that $T$ is the true structure of the simulated network based on SBMs, and $S$ is the analogy from our clustering algorithm, we consider two $n \times n$ binary matrices, $\left(M^{(T)}_{ij}\right)$ and $\left(M^{(S)}_{ij}\right)$, defined respectively as follows:
$$M_{ij}^{(T)} = \begin{cases}
1, \qquad &\mbox{Nodes $i$ and $j$ are in the same community in model $T$},
\\ 0, \qquad &\mbox{otherwise};
\end{cases}$$
and 
$$M_{ij}^{(S)} = \begin{cases}
1, \qquad &\mbox{Nodes $i$ and $j$ are in the same community by algorithm $S$},
\\ 0, \qquad &\mbox{otherwise}.
\end{cases}$$
Our measure is defined on a cluster level, and in the form of an $h \times h$ matrix. The fundamental principle of our measure is simple; that is, we compute the proportion of mis-clustered nodes, from cluster to cluster. More precisely, for all nodes in different communities $k$ and $l$ ($k \neq l$) under $T$, the error measure is given by
$$\epsilon_{kl} = \sum_{\tiny \substack{i, j \mbox{ s.t.} \\ c^{(T)}(i) = k, c^{(T)}(j) = l}} \frac{\left|M_{ij}^{(T)} - M_{ij}^{(S)}\right|}{n_k n_l},$$
where $n_k$ and $n_l$ are the sizes of communities $k$ and $l$, respectively. The error measure for a single community $k$ is defined analogously, i.e.,
$$\epsilon_{kk} = \sum_{\tiny \substack{i, j \mbox{ s.t.} \\ c^{(T)}(i) = c^{(T)}(j) = k}} \frac{\left|M_{ij}^{(T)} - M_{ij}^{(S)}\right|}{n_k (n_k - 1)}.$$  
Ultimately, the error matrix is given by
$$\veps = (\epsilon_{kl})_{h \times h}.$$
We would like to point out that our error matrix measure can be more appealing in some comprehensive analysis, as it shows where (i.e., in which community or communities) exactly a large amount of mis-clustering occurs when the clustering result far deviates from ground truth.

\subsection{Effect of threshold $p$}

In the first experiment, we show that a correct choice of threshold $p$ has a significant impact on the $p$-clique index clustering algorithm (c.f.\ Algorithm~\ref{Alg:bipar}). We simulate an SBM consisting of $120$ nodes which are clustered into three communities of sizes $100$, $10$ and $10$, respectively. The parameters of in-group and cross-group link densities are summarized in Table~\ref{Table:parameterone}.
\begin{table}[ht]
	\renewcommand{\arraystretch}{2}
	\begin{tabular}{|c|c|c|c|}
		\hline
		& Cluster (size: 100) & Cluster 2 (size: 10) & Cluster 3 (size: 10)
		\\ \hline
		Cluster 1 (size: 100) & 0.2 & 0.05 & 0.05
		\\ \hline
		Cluster 2 (size: 10) & 0.05 & 0.5 & 0.05
		\\ \hline
		Cluster 3 (size: 10) & 0.05 & 0.05 & 0.5
		\\ \hline
	\end{tabular}
	\caption{Link densities of clusters of a simulated SBM1}
	\label{Table:parameterone}
\end{table}

The expected clique score of the simulated SBM1 is approximately $0.1597$, which can be used as an estimate of the overall link density of the network. Let the error parameter $\alpha$ be equal to $0.025$. According to Equation~(\ref{Eq:pupper}), the upper bound for the threshold $p$ is $0.09541$. On the other hand, the lower bound for $p$ (c.f.\ Equation~(\ref{Eq:plower})) is $\max\{0.0635, 0.0927\} = 0.0927$. Suppose that we choose $p = 0.09$, simulate 100 independent SBM1s, and compute the distance-based error matrix $\veps$. Albeit the threshold $p$ is just slightly less than the lower bound, we have a large under-clustering error for clusters 2 and 3, i.e., $\epsilon_{23} = \epsilon_{32} = 0.2218$ with standard error ${\rm SE}(\epsilon_{23}) = {\rm SE}(\epsilon_{32}) = 0.0376$, indicating that the probability of misclassifying nodes from these two clusters is about $22\%$ in average. Suppose that the value of $p$ is increased to $0.11$ (even though $p = 0.11$ is greater than the upper bound), the errors $\epsilon_{23} = \epsilon_{32}$ drop dramatically to $0.0614$ with standard error ${\rm SE}(\epsilon_{23}) = {\rm SE}(\epsilon_{32}) = 0.0172$. As $\epsilon_{23} = \epsilon_{32} \approx 6 \%$ is the largest entry in the error matrix $\veps$ for $p = 0.11$, it seems that over-clustering does not bring too much trouble in this experiment; the reason is that all predefined in-group densities are significantly larger than cross-group densities. Due to the limit of space, we refer the interested readers to~\cite[Sections 4.2 and 4.3]{Ouyang} for more analogous examples.

\begin{figure}[ht]
	\begin{center}
		\includegraphics[scale = 0.63]{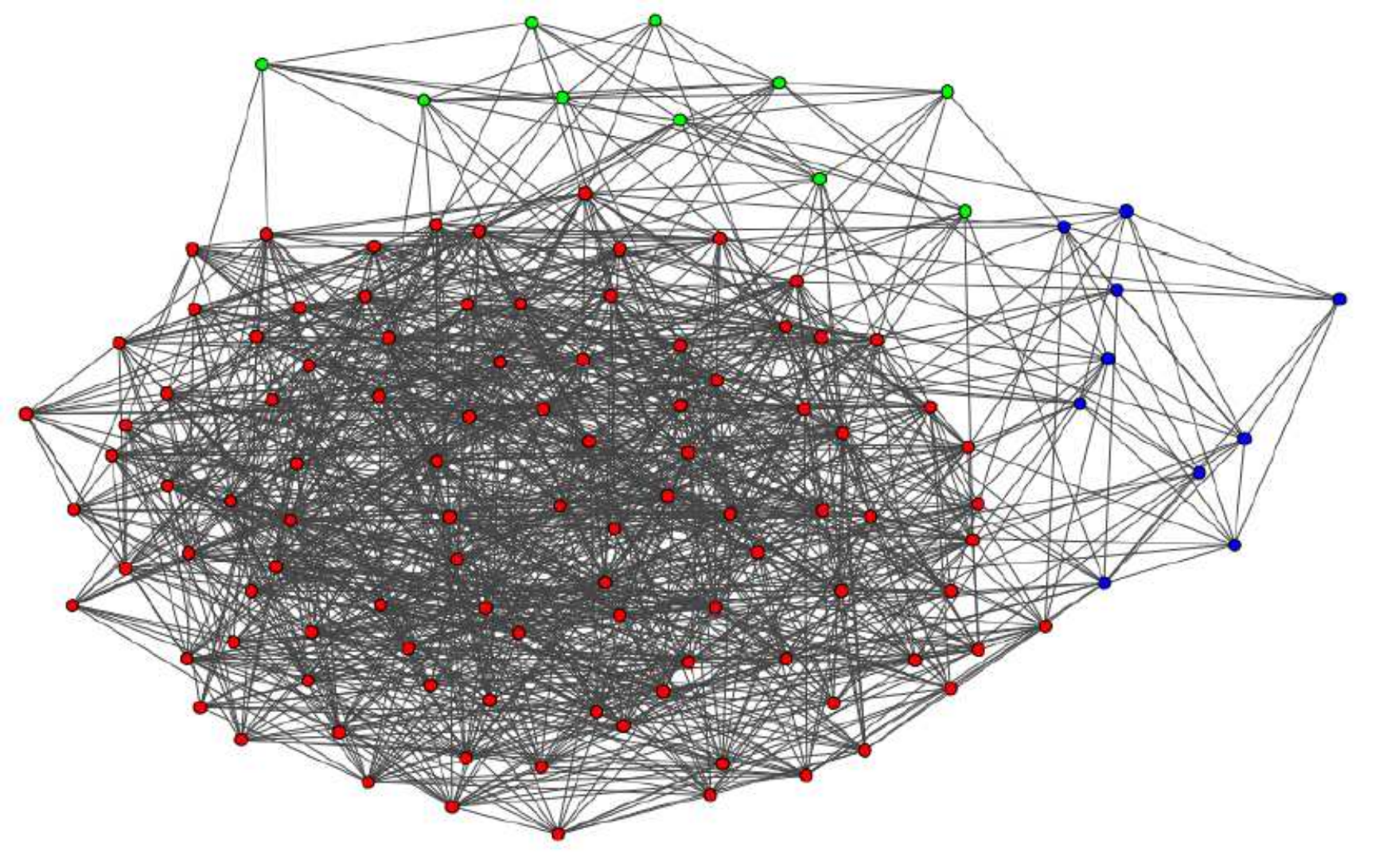}
	\end{center}
	\caption{Clustering result of SBM1 by Algorithm~\ref{Alg:localbipar}; $p = 0.9527$}
	\label{Fig:SBMp}
\end{figure}

Nevertheless, as long as the parameter $p$ is fairly close to the proposed threshold selection interval from the above, the clustering outcomes for this experiment (c.f.\ Table~\ref{Table:parameterone}) are under satisfactory. We choose the optimal value of $p = 0.9527$, and depict the result in Figure~\ref{Fig:SBMp}, where the three communities are clearly identified and colored by blue, red and green.

\subsection{Global threshold v.s.\ localized thresholds}

The next experiment is designed to compare the performance of the algorithms proposed in this manuscript. We show that the localized algorithm provides more reliable outcomes when the threshold $p$ is unknown. We consider SBM2 with $140$ nodes which are clustered into three communities of sizes $100$, $20$ and $20$, respectively. The parameters of in-group and cross-group link densities are summarized in Table~\ref{Table:parameter}.
\begin{table}[ht]
	\renewcommand{\arraystretch}{2}
	\begin{tabular}{|c|c|c|c|}
		\hline
		& Cluster 1 (size: 120) & Cluster 2 (size: 20) & Cluster 3 (size:20)
		\\ \hline
		Cluster 1 (size: 120) & 0.2 & 0.05 & 0.05
		\\ \hline
		Cluster 2 (size: 20) & 0.05 & 0.6 & 0.12
		\\ \hline
		Cluster 3 (size: 20) & 0.05 & 0.12 & 0.8
		\\ \hline
	\end{tabular}
	\caption{Link densities of clusters of simulated SBM2}
	\label{Table:parameter}
\end{table}

The expected clique score of the simulated SBM2 is approximately $0.1546$. We again set the error parameter $\alpha$ at $0.025$. The associated threshold parameter $p$ according to Equation~(\ref{Eq:pupper}) is $0.0959$. As the parameter $p$ is less than the cross-group link density between communities 2 and 3, it seems to be difficult to separate the two clusters with the overall threshold $p$ via Algorithm~\ref{Alg:bipar}. To verify our conjecture, we simulate 100 independent SBM2s, compute the {\rm NMI} for each replication, and take the average as an estimate. We obtain $\widehat{\rm NMI} = 0.8488$ with standard error $0.0025$. In addition, we compute the proposed block-wise distance-based measure, summarized in Table~\ref{Table:error}.
\begin{table}[ht]
	\renewcommand{\arraystretch}{2}
	\begin{tabular}{|c|c|c|c|}
		\hline
		& Cluster 1 & Cluster 2 & Cluster 3
		\\ \hline
		\multirow{2}{*}{Cluster 1} & $\epsilon_{11} = 0.0006$ & $\epsilon_{12} = 0.0.0012$ & $\epsilon_{13} = 0.0012$
		\\ \cline{2-4}
		&${\rm SE}(\epsilon_{11}) = 0.0011$ & ${\rm SE}(\epsilon_{12}) = 0.0003$ & ${\rm SE}(\epsilon_{13}) = 0.0003$
		\\ \hline
		\multirow{2}{*}{Cluster 2} & $\epsilon_{21} = 0.0012$ & $\epsilon_{22} < 10^{-4}$ & $\epsilon_{23} = 0.9890$
		\\ \cline{2-4}
		&${\rm SE}(\epsilon_{21}) = 0.0030$ & ${\rm SE}(\epsilon_{22}) < 10^{-4}$ & ${\rm SE}(\epsilon_{23}) = 0.0010$
		\\ \hline
		\multirow{2}{*}{Cluster 3} & $\epsilon_{31} = 0.0012$ & $\epsilon_{32} = 0.9890$ & $\epsilon_{33} = 0.0010$
		\\ \cline{2-4}
		&${\rm SE}(\epsilon_{31}) = 0.0030$ & ${\rm SE}(\epsilon_{32}) = 0.0010$ & ${\rm SE}(\epsilon_{23}) = 0.0009$
		\\ \hline
	\end{tabular}
	\caption{The block-wise error measure of clustering SBM2 via Algorithm~\ref{Alg:bipar}}
	\label{Table:error}
\end{table}

\begin{figure}[ht]
	\begin{center}
		\includegraphics[scale = 0.63]{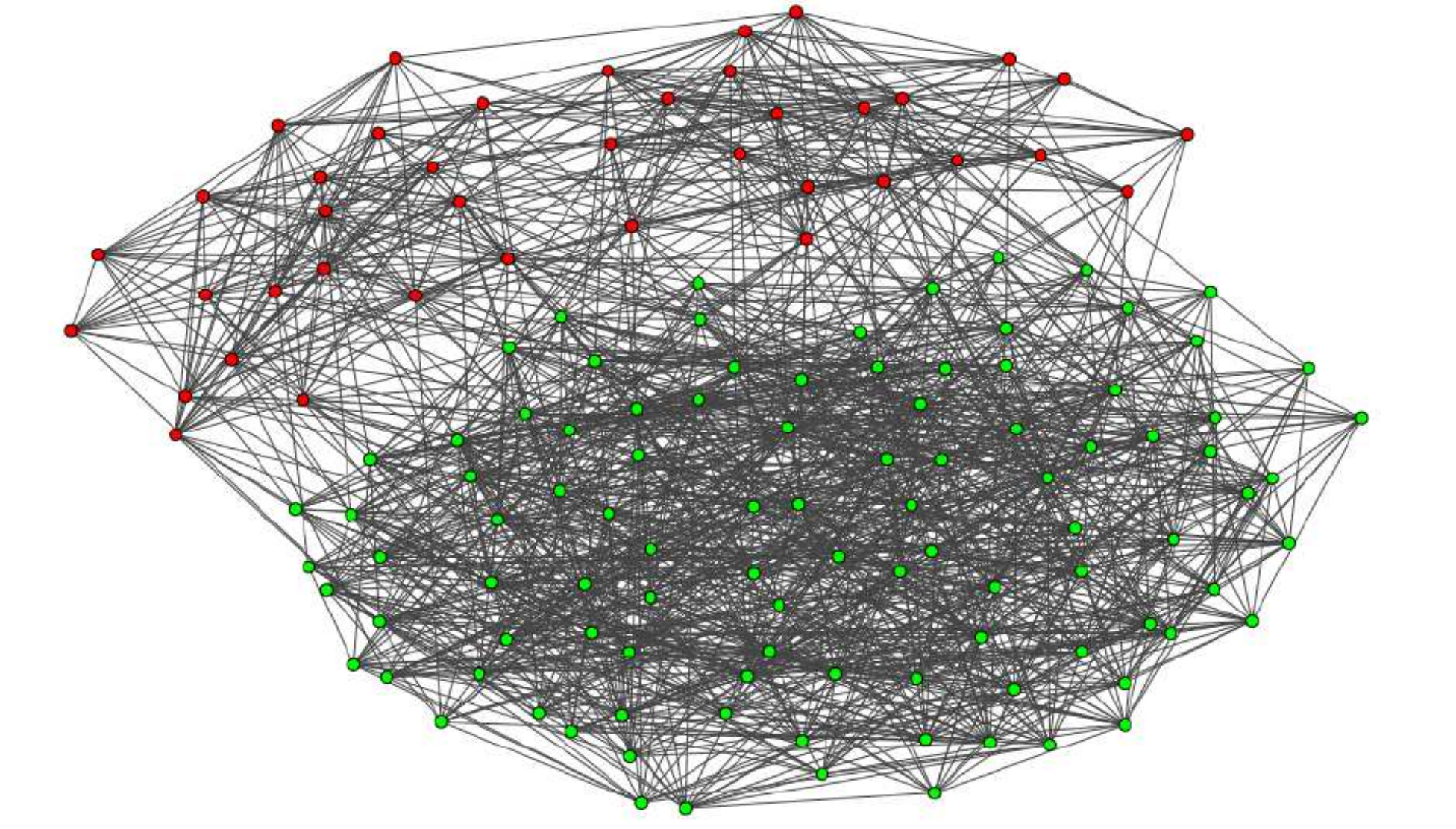}
	\end{center}
	\caption{Clustering result of SBM2 by Algorithm~\ref{Alg:bipar}}
	\label{Fig:SBMbipar}
\end{figure}

Although community $1$ is successfully identified, we observe that the estimate of error rate between communities $2$ and $3$ is $\epsilon_{23} = 0.9890 = 98.90 \%$, which suggests that the {\rm BiPartition} algorithm with the global threshold $p = 0.0959$ fails to separate these two communities almost surely. This is also reflected in the clustering result (via Algorithm~\ref{Alg:bipar}) of the simulated SBM (c.f.\ Table~\ref{Table:parameter}) given in Figure~\ref{Fig:SBMbipar}. The entire network is divided into two cluster (rather than three), which are colored by red and green.

We remedy this problem by applying the localized clustering algorithm, i.e., Algorithm~\ref{Alg:localbipar}. We start with the root node (the original simulated SBM) in the binary tree structure. We use threshold $p = 0.0959$ to bipartition the root node, and obtain community $1$, and subnetwork $2$, where the latter requires for further clustering. Subnetwork $2$ consists of $40$ nodes and the expected clique score is approximately $0.4026$. With the same $\alpha = 0.025$, the updated threshold parameter for subnetwork $2$ is $0.2536$. We would like to mention that in practice we also compute the threshold parameter for subnetwork $1$ (community $1$), and obtain $0.1231$ for implementing Algorithm~\ref{Alg:localbipar}, where we find that no further subdivision is needed. After subdividing subnetwork $2$, we continue to compute new thresholds for both subsequent subnetworks, and ultimately find that they form the other two communities in the network.

\begin{figure}[ht]
	\begin{center}
		\includegraphics[scale = 0.63]{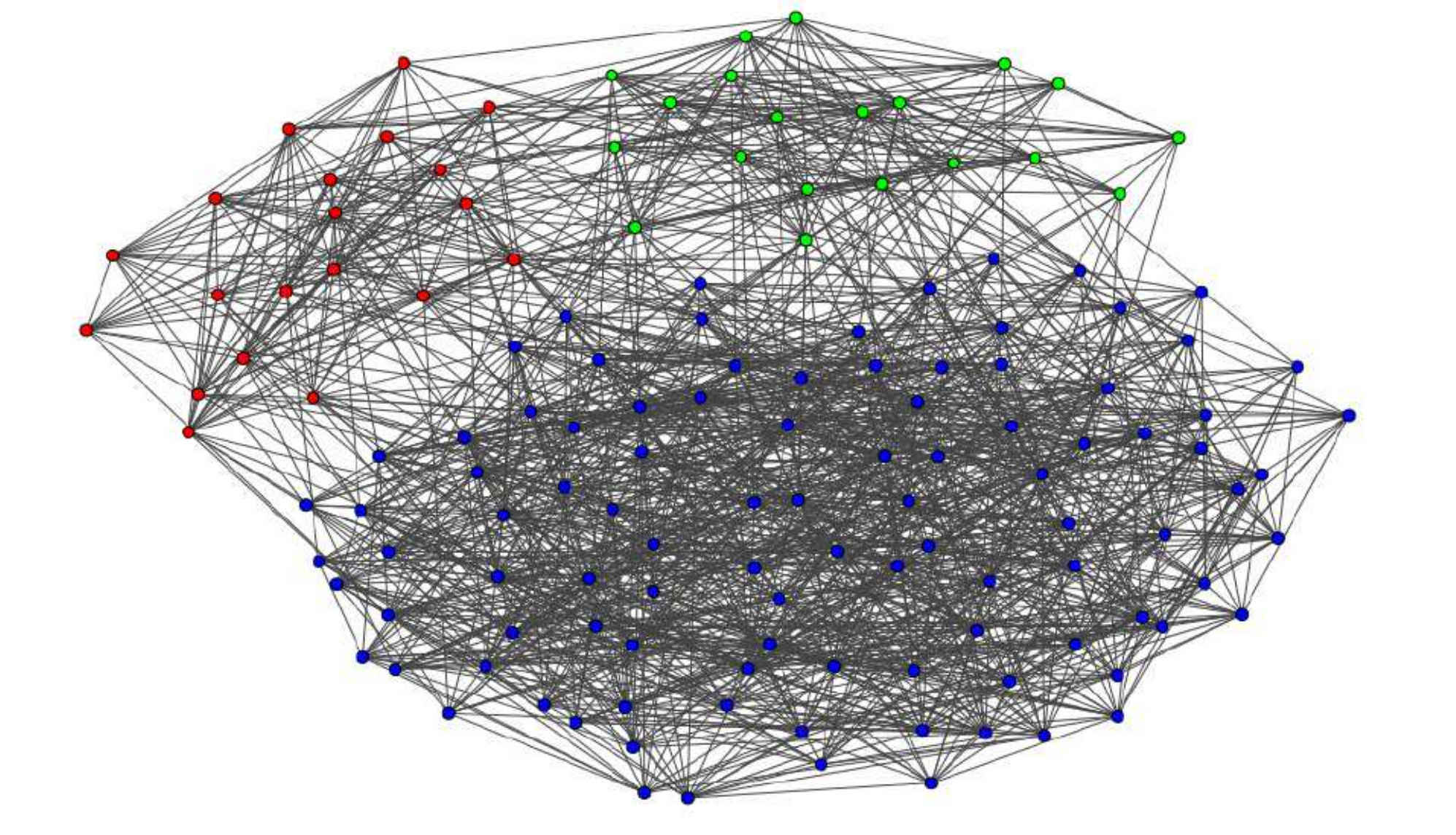}
	\end{center}
	\caption{Clustering result of SBM2 by Algorithm~\ref{Alg:localbipar}}
	\label{Fig:SBMbiparlocal}
\end{figure}

Another $100$ independent SBMs are simulated, and the {\rm NMI} for each simulated network is computed. The mean estimate is $\widehat{\rm NMI}_{\rm local} = 0.9677$ with standard error $0.0034$. We also compute the block-wise error rates, and present them in Table~\ref{Table:errorlocal}, where we find a significant improvement of $\epsilon_{23} = \epsilon_{32} = 0.0019$ versus $0.9890$. The small error rate implies that the communities $2$ and $3$ are significantly identifiable according to the updated threshold for subnetwork $2$. Analogously, we depict the clustering result of a simulated SBM via Algorithm~\ref{Alg:localbipar}, shown in Figure~\ref{Fig:SBMbiparlocal}. The network is successfully divided into three communities as predefined in Table~\ref{Table:parameter}. The communities are colored by blue, red and green.
\begin{table}[ht]
	\renewcommand{\arraystretch}{2}
	\begin{tabular}{|c|c|c|c|}
		\hline
		& Cluster 1 & Cluster 2 & Cluster 3
		\\ \hline 
		\multirow{2}{*}{Cluster 1} & $\epsilon_{11} = 0.0231$ & $\epsilon_{12} < 10^{-4}$ & $\epsilon_{13} = 0.0001$
		\\ \cline{2-4}
		&${\rm SE}(\epsilon_{11}) = 0.0033$ & ${\rm SE}(\epsilon_{12}) < 10^{-4}$ & ${\rm SE}(\epsilon_{13}) < 10^{-4}$
		\\ \hline
		\multirow{2}{*}{Cluster 2} & $\epsilon_{21} < 10^{-4}$ & $\epsilon_{22} = 0.0334$ & $\epsilon_{23} = 0.0019$
		\\ \cline{2-4}
		&${\rm SE}(\epsilon_{21}) < 10^{-4}$ & ${\rm SE}(\epsilon_{22}) = 0.0057$ & ${\rm SE}(\epsilon_{23}) = 0.0019$
		\\ \hline
		\multirow{2}{*}{Cluster 3} & $\epsilon_{31} < 10^{-4}$ & $\epsilon_{32} = 0.0019$ & $\epsilon_{33} = 0.0357$
		\\ \cline{2-4}
		&${\rm SE}(\epsilon_{31}) < 10^{-4}$ & ${\rm SE}(\epsilon_{32}) = 0.0019$ & ${\rm SE}(\epsilon_{23}) = 0.0067$
		\\ \hline
	\end{tabular}
	\caption{The block-wise error measure of clustering SBM2 via Algorithm~\ref{Alg:localbipar}}
	\label{Table:errorlocal}
\end{table}

\subsection{The propose algorithm v.s. modularity maximization}

Notice that the algorithm proposed in this manuscript is inspired from the modularity maximization algorithm. In this section, we would like to compare the proposed algorithm with the modularity approach. In the literature, there is an extensive discussion about the limitations and drawbacks of the modularity-based algorithms. One of the most significant problems of the modularity method is that the maximization of modularity score is not always consistent with the optimized clustering outcome, which was addressed by~\cite{Fortunato}. This limitation is related to under-clustering, as it seems that the modularity maximization algorithm intends to merge small clusters, especially in large-size networks. It is believed that the modularity-based methods undergo over-clustering problems as well. Both of these issues were stressed and discussed in~\cite{Reichardt} and~\cite{Lancichinetti}. To the best of our knowledge, neither of them has been systematically solved up to date. We refer the readers to~\cite{Bickle} for extensive discussions about inconsistency of clustering results from the modularity maximization algorithm. On the other hand, both over-clustering and under-clustering concerns are considered and well controlled in our algorithm, which empirically ensures consistency. We present the following three simulation examples to show resistance of the proposed algorithm in this manuscript.

We reconsider SBM1, a simulated network contains two extremely small communities which are extremely loosely linked; SBM2, a simulated network contains two relatively small communities which are fairly loosely linked; and SBM3, a simulated network which is partitioned into communities of approximately even size. The structural parameters are summarized in Table~\ref{Table:parameterthree}.

\begin{table}[ht]
	\renewcommand{\arraystretch}{2}
	\begin{tabular}{|c|c|c|c|}
		\hline
		& Cluster 1 (size: 60) & Cluster 2 (size: 40) & Cluster 3 (size:40)
		\\ \hline
		Cluster 1 (size: 60) & 0.2 & 0.05 & 0.05
		\\ \hline
		Cluster 2 (size: 40) & 0.05 & 0.6 & 0.12
		\\ \hline
		Cluster 3 (size: 40) & 0.05 & 0.12 & 0.8
		\\ \hline
	\end{tabular}
	\caption{Link densities of clusters of a simulated SBM3}
	\label{Table:parameterthree}
\end{table}

We respectively apply the modularity maximization algorithm and the localized clique-based algorithm to these three networks, and evaluate the clustering results by using the mean estimates of both NMI and ARI, based on 100 independent copies of each simulated network. The results are presented in Table~\ref{Table:modelcompare}.

\begin{table}[ht]
	\renewcommand{\arraystretch}{2}
	\begin{tabular}{|c|c|c|c|c|c|c|c|c|}
		\hline
		\multirow{2}{*}{Network} & \multicolumn{4}{|c|}{Modularity maximization} & \multicolumn{4}{|c|}{Localized clique-based method}
		\\ \cline{2-9}
		& $\widehat{{\rm NMI}}$ & ${\rm SE}(\widehat{{\rm NMI}})$ & $\widehat{{\rm ARI}}$ & ${\rm SE}(\widehat{{\rm ARI}})$ 	& $\widehat{{\rm NMI}}$ & ${\rm SE}(\widehat{{\rm NMI}})$ & $\widehat{{\rm ARI}}$ & ${\rm SE}(\widehat{{\rm ARI}})$
		\\ \hline
		SBM1 & 0.1681 & 0.0057 & 0.0514 & 0.0056 & 0.9309 & 0.0047 & 0.8556 & 0.0074
		\\ \hline
		SBM2 & 0.7525 & 0.0092 & 0.5159 & 0.0106 & 0.9677 & 0.0030 & 0.9525 & 0.0035
		\\ \hline
		SBM3 & 0.9568 & 0.0048 & 0.9055 & 0.0054 & 0.9477 & 0.0039 & 0.8719 & 0.0064
		\\ \hline
	\end{tabular}
	\caption{Comparison of clustering results respectively from the modularity maximization algorithm and the localized clique-based algorithm}
	\label{Table:modelcompare}
\end{table}

Based on the experimental results, we observe that the modularity maximization algorithm performs extremely poorly for SBM1. As a matter of fact, the modularity algorithm even fails to distinguish the two smaller communities (c.f.\ clusters 2 and 3). As the sizes of small communities and the link density there-between increase, the modularity maximization algorithm recovers (as reflected in the results for SBM2), but apparently still does not appear as well-performed as our algorithm. For networks comprising similar-size communities (e.g., SBM3), the modularity maximization algorithm seems to outperform our algorithm, but not significantly. Having the randomness of simulated networks in mind, we thus conclude that the proposed algorithm in this manuscript is preferred due to its consistency and reliability.

\subsection{Time complexity}
In the era of big data, researchers always concern about the efficiency of algorithms, especially the newly proposed algorithm, when the network size is large. In the last part of this section, we look into this issue numerically by implementing the localized algorithm in {\rm Python} for several networks of different sizes and structure. Both mean estimates of NMI and ARI are computed. Our testing results are summarized in Table~\ref{Table:time}. We are convinced that the proposed algorithm is able to provide accurate clustering results in a relatively short amount of time.
\begin{table}[ht]
	\renewcommand{\arraystretch}{2}
	\begin{tabular}{|c|c|c|c|c|c|}
		\hline
		Size & Number of clusters & Number of simulations & $\widehat{\rm NMI}$ & $\widehat{\rm ARI}$ & Time
		\\ \hline
		120 & 3 & 100 & 0.8596 & 0.9309 & 92.7 ms
		\\ \hline 1270 & 4 & 100 & 0.9687 & 0.9779 & 94.5 ms
		\\ \hline 7000 & 10 & 20 & 0.9895 & 0.9901 & 1.66 s 
		\\ \hline 20000 & 25 & 20 & 0.8960 & 0.8869 & 12.6 s
		\\ \hline
	\end{tabular}
	\caption{A summary of community detection for several simulated SBMs via Algorithm~\ref{Alg:localbipar}}
	\label{Table:time}
\end{table} 

\section{Concluding remarks}
\label{Sec:con}
In this section, we address several remarks and discuss some possible future work. To conclude, we propose a new clique-based measure in this manuscript to evaluate the quality of network clustering, and design an algorithm to maximize an objective function based on the proposed measure. The clique score of each community in our clustering result is guaranteed to be higher than some predetermined threshold $p$. We also consider the situation at which the threshold $p$ is unspecified. We develop an approach which accounts for over-clustering and under-clustering problems simultaneously to select localized $p$ for the network and its subsequent networks. An associated localized algorithm is proposed and discussed.

Studies of networks or network models usually involve big data problems. When network size or parameter space or both are large, it is always challenging to use model-based clustering methods, as many of them depend on accurate but slow Bayesian MCMC algorithms, for example~\cite{Snijders}. However, the methods proposed in this manuscript attempt to convert clustering analysis to optimization problems. Therefore, many sophisticated machine learning techniques and well-developed approximation methods are ready to use to deal with big data issues.

It is worthy of mentioning that the methods considered in this manuscript also applicable to sparse social network, since the computation of $p$-clique index is primarily based on the $p$-clique matrix $\vC(p) = \vA - p(\vone_{n \times n} - \vI_{n \times n})$. This matrix is not sparse even if the adjacency matrix $\vA$ is sparse. Numerical methods, such as the Lanczos method in~\cite{Calvetti}, promise that the eigenvector corresponding to the largest eigenvalue of $\vC(p)$ always can be determined very fast.

The development of the localized clustering algorithm in this paper depends on a strong assumption that the null network is an \ER\ model. This may not be true for many real-world networks. It is suggested in~\cite{Barabasi} that many networks around us follow {\em power law}, which is not the case for the \ER\ graphs. We conjecture that it may be more accurate to consider a scale-free network as null for the studies of social networks possessing power-law property, and we will consider future research in this direction. 



\begin{thebibliography}{}
	%
	%
	\bibitem{Airoldi}
	\textsc{Airoldi, E.M., Blei, D.M., Fienberg, S.E.} and \textsc{Xing, E.P.}
	Mixed membership stochastic blockmodels. 
	{\it Journal of Machine Learning Research}, {\bf 9}, 1981--2014. (2008)
	
	\bibitem{Barabasi}
	\textsc{Barab\'{a}si, A.-L.} and \textsc{Albert, R.} Emergence of scaling in random networks. {\em Science}, {\bf 286}, 509--512. (1999)
	
	\bibitem{Bickle}
	\textsc{Bickel, P.J.} and \textsc{Chen, A.} A nonparametric view of network models and Newman-Girvan and other modularities {\it Proceedings of the National Academy of Sciences of the United States of America}, {\bf 106}, 21068--21073. (2009)
	
	\bibitem{Calvetti}
	\textsc{Calvetti, D., Reichel, L.} and \textsc{Sorensen, D.}
	An implicitly restarted Lanczos method for large symmetric eigenvalue problems. {\it Electronic Transactions
		on Numerical Analysis}, {\bf 2}, 1--21. (1994)
	
	\bibitem{Chung}
	\textsc{Chung, F.R.K.} {\it Spectral Graph Theory}. American Mathematical Society, Providence, RI. (1997)
	
	\bibitem{Erdos}
	\textsc{Erd\"{o}s, P.} and \textsc{R\'{e}nyi, A.} On random graphs {\rm I}. {\it Publicationes Mathematicae}, {\bf 6}, 290--297. (1959)
	
	\bibitem{Fortunato}
	\textsc{Fortunato, S.} and \textsc{Barth\'{e}lemy, M.} Resolution limit in community detection. {\it Proceedings of the National Academy of Sciences of the United States of America}, {\bf 104}, 36--41. (2007)
	
	\bibitem{Fred}
	\textsc{Fred, A.} and \textsc{Jain, A.} Robust data clustering, in the {\em Proceedings of IEEE Computer Society Conference on Computer Vision and Pattern Recognition}, {\bf 2}, 128--133. (2003)
	
	\bibitem{Gilbert}
	\textsc{Gilbert, E. N.} Random graphs. {\it Annals of Mathematical Statistics}, {\bf 30}, 1141--1144. (1959) 
	
	\bibitem{Goldenberg}
	\textsc{Goldenberg, A., Zheng, A.X., Fienberg, S.E.} and \textsc{Airoldi, E.M.}  A survey of statistical network models. {\it Foundations and Trends in Machine Learning}, {\bf 2}, 129--233. (2010)
	
	\bibitem{Handcock}
	\textsc{Handcock, M.S., Raftery, A.E.} and \textsc{Tantrum, J.M.} Model-based clustering for social networks. 
	{\it Journal of the Royal Statistical Society, Series A}, {\bf 170}, 301--354. (2007)
	
	\bibitem{Hoff}
	\textsc{Hoff, P.D., Raftery, A.E.} and \textsc{Handcock, M.S.} Latent space approaches to social network analysis. 
	{\it Journal of the American Statistical Association}, {\bf 97}, 1090--1098. (2002)
	
	\bibitem{Holland1981}
	\textsc{Holland, P. W.} and \textsc{Leinhardt, S.} An exponential family of probability distributions for directed graphs. {\it Journal of the American Statistical Association}, {\bf 76}, 33--50. (1981)
	
	\bibitem{Holland1983}
	\textsc{Holland, P.W., Laskey, K.B.} and \textsc{Leinhardt, S.} Stochastic blockmodels: First steps. {\it Social Networks}, {\bf 5}, 109--137. (1983)
	
	\bibitem{Horn}
	\textsc{Horn, R.A.} and \textsc{Johnson, C.R.} {\it Matrix Analysis}. Cambridge University Press, New York. (1985)
	
	\bibitem{Hubert}
	\textsc{Hubert, L.} and \textsc{Abrabie, P.} Comparing partitions. {\it Journal of Classification}, {\bf 2}, 193--218. (1985
	
	\bibitem{Lancichinetti}
	\textsc{Lancichinetti, A.} and \textsc{Fortunato, S.} Limits of modularity maximization in community detection. {\it Physical Review E}, {\bf 84}, 066122. (2011)
	
	\bibitem{Newman2001a}
	\textsc{Newman, M.E.J.} The structure of scientific collaboration networks. {\it Proceedings of the National Academy of Sciences of the United States of America}, {\bf 98}, 404--409. (2001)
	
	\bibitem{Newman2001b}
	\textsc{Newman, M.E.J., Strogatz, S.H.} and \textsc{Watts, D.J.}. Random graphs with arbitrary degree distributions and their applications. {\it Physical Review E}, {\bf 64}, 026118. (2001)
	
	\bibitem{Newman2006}
	\textsc{Newman, M.E.J.} Modularity and community structure in networks. {\it Proceedings of the National Academy of Sciences of the United States of America}, {\bf 103}, 8577--8582. (2006)
	
	\bibitem{Ng}
	\textsc{Ng, A.Y., Jordan, M.I.} and \textsc{Weiss, Y.} On spectral clustering: Analysis and an algorithm. {\it Advances in Neural Information Processing Systems}, {\bf 14}, 849--856. (2001)
	
	\bibitem{Snijders}
	\textsc{Snijders, T.A.B.} and \textsc{Nowicki, K.}
	Estimation and prediction for stochastic blockmodels
	for graphs with latent block structure. {\it Journal of Classification}, {\bf 14}, 75--100. (1997)
	
	\bibitem{Ouyang}
	\textsc{Ouyang, G.}
	{\em Social Network Community Detection}. Ph.D.\ dissertation.
	University of Connecticut. (2015)
	
	\bibitem{Pao}
	\textsc{Pao, L.-F.} Discovering the dynamics of smart business networks. {\it Computational Management Science}, {\bf 1}, 445--458. (2014)
	
	\bibitem{Pei}
	\textsc{Pei, X., Zhan, X.-X.} and \textsc{Jin, Z.} Application of pair approximation method to modeling and analysis of a marriage network. {\it Applied Mathematics and Computation}, {\bf 294}, 280--293. (2017)
	
	\bibitem{Reichardt}
	\textsc{Reichardt, J.} and \textsc{Bornholdt, S.} Statistical mechanics of community detection. {\it Physical Review E}, {\bf 74}, 016110. (2006)
	
	\bibitem{Shi}
	\textsc{Shi, J.} and \textsc{Malik, J.} Normalized cuts and image segmentation. {\it IEEE Transaction on Pattern Analysis and Machine Intelligence}, {\bf 22}, 888--905. (2000)
	
	\bibitem{Watts}
	\textsc{Watts, D.J.} and \textsc{Strogatz, S.H.} Collective dynamics of ``small-world'' networks. {\it Nature}, 440--442. (1998)
	
	\bibitem{Wohlgemuth}
	\textsc{Wohlgemuth, J.} and \textsc{Matache, M.T.} Small-wold properties of Facebook group networks. {\it Complex Systems}, {\bf 23}, 197--225. (2014) 
	
\end{thebibliography}


\end{document}